# High degree of chaos synchronization of a single pair of transverse modes with different polarizations in vector lasers subjected to self-mixing modulation

Kenju Otsuka

*TS³L Research*

*Yamaguchi 126-7, Tokorozawa 359-1145, Japan*

*e-mail address: kenju.otsuka@gmail.com*

Synchronization of chaos among a single pair of transverse modes featuring intermode dynamical non-independence, resulting from cross-saturation of modal population inversions and coherent modal field coupling, is explored in a thin-slice solid-state laser with coupled orthogonally polarized transverse modes operating in quasi-locked states. The chaotic signals that are created in one of the pairing modes (sender) by self-mixing modulations are shown to be transferred solely to the partner mode (receiver) among all the other modal fields with different polarizations. The experimental results are reproduced in a simulation of a model equation of coupled laser modes subjected to self-mixing modulations.

Synchronization phenomena of chaotic oscillators [1] are encountered in physical, chemical, and biological systems, as well as in laser systems. In the context of coherently coupled chaotic oscillators, there are a number of different interpretations of chaos synchronization, such as master-slave synchronization and synchronization based on mutually coupled oscillators. In laser physics, a variety of chaos synchronizations have been reported in *coherent* coupling of solid-state lasers [2, 3] and semiconductor lasers [4, 5]. Besides, polarization synchronization was reported in unidirectionally coupled vertical-cavity surface-emitting semiconductor lasers, where chaos synchronization occurred when the polarization of the master laser was perpendicular to that of the free-running slave laser [6, 7].

Different forms of chaos synchronization owing to *incoherent* coupling through cross-saturation of population inversions among modes have been demonstrated in a modulated multimode laser, depending on the pump power [8]. Chaos synchronizations based on the similar cross-saturation mediated *incoherent* coupling of orthogonally polarized modes have been reported in a dual-polarization laser [9].

In this paper, self-organized collective chaos synchronization is reported in the regime of quasi-locked states of different orthogonally polarized transverse eigenmodes in a thin-slice solid-state laser subjected to self-mixing modulations based on the combined effect of *coherent* modal coupling and *incoherent* cross-saturation dynamics among modes. A high degree of chaos synchronization, with amplitude correlation coefficients R > 0.99, is found to occur for a particular pair of polarized transverse modal fields, which are embedded in polarization vector fields formed from coupled transverse eigenmodes. Synchronized chaotic relaxation oscillations have been demonstrated using different schemes, i.e., self-mixing modulation of the total output or one of the pairing modes.

The results of the experiments are reproduced in simulations of a model of coupled laser mode equations subjected to self-mixing modulations including the effects of cross-saturation of population inversions and field coupling among oscillating transverse modes.

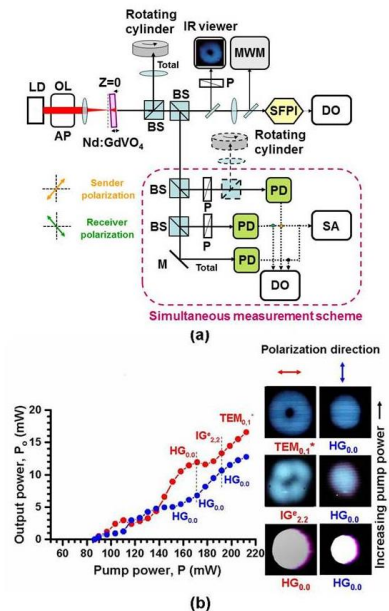

FIG. 1. (a) Experimental apparatus. LD: laser diode, AP: anamorphic prism pair, OL: objective lens, BS: beam splitter, M: mirror, P: polarizer, MWM: multi-wavelength meter, SFPI: scanning Fabry-Perot interferometer, PD: photo-diode, DO: digital oscilloscope, SA: spectrum analyzer. (b) Input-output characteristics and pump-dependent DPO far-field intensity patterns.

The experimental setup is shown in Fig. 1(a). A nearly collimated lasing beam from a laser diode (wavelength: 808 nm) was passed through an anamorphic prism pair to transform an elliptical beam into a circular one, and it was focused onto a thin-slice laser crystal by a microscopic objective lens of numerical aperture NA = 0.5. The laser crystal was a 3 mm-diameter clear-aperture, 1 mm-thick, 3 at%-doped c-cut Nd:GdVO$_4$ whose end surfaces were directly coated with dielectric mirrors M$_1$ (transmission at 808 nm > 95%; reflectance at 1064 nm = 99.8%) and M$_2$ (reflectance at 1064 nm = 99%). Lasing optical spectra were measured by a multi-wavelength meter (HP-86120B; wavelength range, 700–1650 nm) for obtaining global views and a scanning Fabry-Perot interferometer (Burleigh SA$^{PLUS}$; 2 GHz free spectral range; 6.6 MHz resolution) for measuring detailed structures. The polarization resolved waveform measurement scheme is boxed by dashed line.

In the case of thin-slice solid-state lasers with coated end mirrors, abbreviated as TS$^3$Ls hereafter, a stable resonator condition is achieved through the thermally induced lensing effect [10], and the input-output characteristics as well as the transverse and longitudinal mode oscillation properties depend directly on the focusing condition (e.g., spot size and shape) of the pump beam on the crystal due to the mode-matching between the pump and lasing mode profiles [11, 12]. In the experiment, the pump-beam diameter was changed by shifting the laser crystal along the z-axis, as depicted in Fig. 1(a). The pump spot size, $w_p$, increased as the laser crystal was shifted away from the pump-beam focus along the z-axis (i.e., z > 0). When $w_p$ exceeded about 80 μm, dual-polarization oscillations (DPOs) were observed, starting from TEM$_{00}$ at the threshold and leading to various transverse modes. When the pump position was precisely changed by moving the laser sample along the x-axis or y-axis with an accuracy of 10 μm and a small tilt of the 1-mm-thick cavity of $|a| \leq 1.5°$ was made with an accuracy of 0.3° as depicted in the inset, several transverse modes appeared at fixed $w_p$.

Typical input-output characteristics are shown in Fig. 1(b). Here, $w_p$ = 80 μm, DPO starts at a threshold pump power of $P_{th}$ = 85 mW in the Hermite-Gaussian HG$_{0,0}$ modes for orthogonal polarizations. At pump powers P greater than 170 mW, the horizontally polarized mode undergoes successive structural changes, featuring the Ince-Gaussian IG$^e_{2,2}$ mode and 'doughnut' mode, while the vertically polarized HG$_{0,0}$ mode is preserved. The DPO characteristics shown in Fig. 1(b) were reproducibly obtained by adjusting the crystal and pump positions [x, y, z] as well as the tilt angle. Assume that the transverse mode locking occurs in the form of **u** = **u$_1$** + r e$^{i\Delta\phi}$ **u$_2$**, where **u$_1$**, **u$_2$** are eigenfuctions of pump-dependent DPO eigenmodes shown in Fig. 2(b), r = $(I_2/I_1)^{1/2}$ is the modal field amplitude ratio and $\Delta\phi$ is their phase difference [13]. Although numerically reconstructed polarization-resolved patterns were resemble to the experimental results, complete transverse mode locking was not established as will be discussed below.

The output DPO laser beam was passed through a polarizer, and the polarization-dependent changes in the far-field patterns were observed. Experimental results corresponding to the orthogonally polarized HG$_{0,0}$ pair are shown in Fig. 2(a). The global oscillation spectrum measured by the multi-wavelength meter indicated a single longitudinal mode at a wavelength of λ = 1065.58 nm, while the adjacent longitudinal mode separated by $\Delta\lambda = \lambda^2/2nL$ = 0.258 nm was not observed (n: refractive index).

Detailed optical spectra corresponding to orthogonally polarized HG$_{0,0}$ mode operation, as measured by the SFPI, are shown on the left of Fig. 2(b). Apart from the complete transverse mode locking of orthogonally polarized modes, as demonstrated in a c-cut Nd:GdVO$_4$ laser with the semi-confocal external cavity, where the pure single-frequency operation occurs [13], peculiar, strongly asymmetric optical mode spectra, each consisting of dominant peaks and weak subsidiary peaks, appear for the orthogonally polarized modes. Despite the degenerate HG$_{0,0}$ modes in the cold cavity, thermal birefringence in c-cut vanadate crystals [14, 15] is evident in TS$^3$L, unlike external cavity lasers whose transverse modes are predominantly determined by the cavity configuration. Moreover, distortion of the transverse mode, presumably due to aberration/astigmatism of a thermal-induced lens, can be seen, e.g., the 'elongated' HG$_{0,0}$ profile in Fig. 2(a). The resulting frequency detuning among orthogonally polarized modes would prevent complete transverse mode locking. In fact, oscillation frequency detuning on the order of 400 MHz remains in the present case.

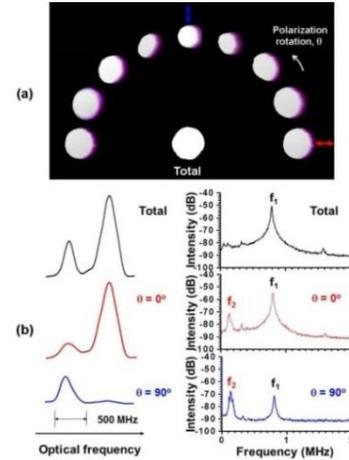

FIG. 2. (a) Polarization-dependent far-field patterns. (b) Optical and power spectra of DPO eigenmodes. P = 172 mW.

However, it will be shown later that the observed orthogonally polarized transverse modes are not independent, and they form partially coherent fields, namely, a quasi-locked state, through mutual nonlinear interaction of closely spaced orthogonally polarized HG$_{0,0}$ modes. The power spectra of the orthogonally polarized eigenmodes are shown on the right of Fig. 2(b), together with that for the total output. The f$_1$-peak corresponds to the fundamental relaxation oscillation noise, while the f$_2$-peak arises through transverse cross-saturation of population inversions of orthogonally polarized modes [16]. Note that the lower frequency peak at $f_2$ is suppressed in the total output. This implies *inherent antiphase dynamics* featuring intermode dynamical non-independence in multimode lasers, where *the total output is self-organized to behave as a single mode laser* that exhibits a noise peak only at $f_1$ and there are no peaks below $f_1$ [17, 18].

Figure 3(a) shows noise intensities at $f_1$ and $f_2$ as a function of the polarizer angle, $\theta$. It should be noted that noise intensities at the lower relaxation oscillation frequencies, $f_2$, which result from the cross-saturation dynamics of modal population inversions, rapidly decrease as $\theta$ approaches two critical angles $\pm\theta_c$, which are symmetric with respect to the x and y axes. The corresponding power spectra are shown in Fig. 3(b). It is apparent that the $f_2$ peak due to the inherent antiphase dynamics is strongly suppressed below the measurement-system noise level for a pair of polarized transverse fields, similar to the case of the total output spectrum in the right column of Fig. 2(b). Complete suppression of the frequency noise peaks below $f_1$ does not occur because of the *incoherent* effect of cross-saturation of population inversions in multimode solid-state lasers, in which *coherent* phase-sensitive interactions of modal fields is absent [18]. Therefore, this phenomenon suggests that such a pair of transverse modes behaves as a coherent single mode, reflecting the influence of transverse mode locking.

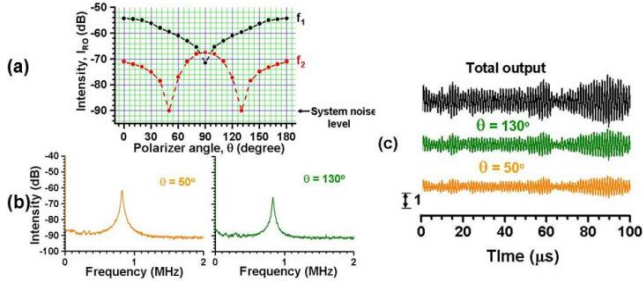

FIG. 3. (a) Polarization-dependent relaxation oscillation intensity. (b) Power spectra for a pair of transvesr modes. (c) Noise-driven modal and total output fluctuations. Pump power, P = 172 mW.

On the other hand, random modal relaxation-oscillation fluctuations of a pair of polarized modal outputs, which are driven by mode-partitioned spontaneous emission noise, and the total output were found to exhibit a high degree of synchronization. Part of the output beam was divided into three beams for simultaneous measurement of the two signals in different polarizations and the total output. It is apparent that noise-driven relaxation oscillations of a pair of polarized modal outputs and the total output are synchronized, where their amplitude correlation coefficient given by $R = \Sigma_i (I_{1,i} - \langle I_1\rangle)(I_{2,i} - \langle I_2\rangle)/ [\Sigma_i (I_{1,i} - \langle I_1\rangle)]^{1/2}[\Sigma_i (I_{1,i} - \langle I_1\rangle)]^{1/2}$ was as large as 0.992. Here, simultaneous measurements were carried out by three InGaAs photodiodes (New Focus 1811, DC-125 MHz) followed by a digital oscilloscope (Tektronics TDS 3052, DC-500 MHz). Zoomed in views of waveforms are shown in Fig. 3(c).

Now, a highly sensitive self-mixing modulation experiment [3, 19] owing to the large fluorescence-to-photon lifetime ratio [20] was carried out to investigate the response of the vector laser, where 50% of the total beam was focused onto a rotating Al-cylinder as shown in Fig. 1(a). Here, the effective interference is expected among each linearly polarized component of randomly polarized scattered fields and the corresponding polarization resolved lasing field pattern with the same polarization [21].

When the Doppler-shift frequency, $f_D = 2v/\lambda$ (v: moving speed along the laser axis), was tuned around the relaxation oscillation frequency, chaotic relaxation oscillations were easily brought about. The antiphase chaotic relaxation oscillations of DPO eigenmodes, together the corresponding power spectra, are shown in Fig. 4(a). The chaotic attractor of each mode inherits the antiphase dynamic character of the corresponding stationary state shown in Fig. 2(b), and the unsynchronized chaotic relaxation oscillations are obvious. While, the generic nature of a single pair of polarized transverse fields is preserved such that the pair behaved as a coherent mode, exhibiting a high degree of chaos synchronization as for a single pair of transverse modes. Results are shown in Fig. 4(b), together with the corresponding power spectrum without $f_2$ peak and correlation plots indicating R = 0.993. Chaos synchronization was readily suppressed when $\theta$ was shifted slightly from $\theta_c$.

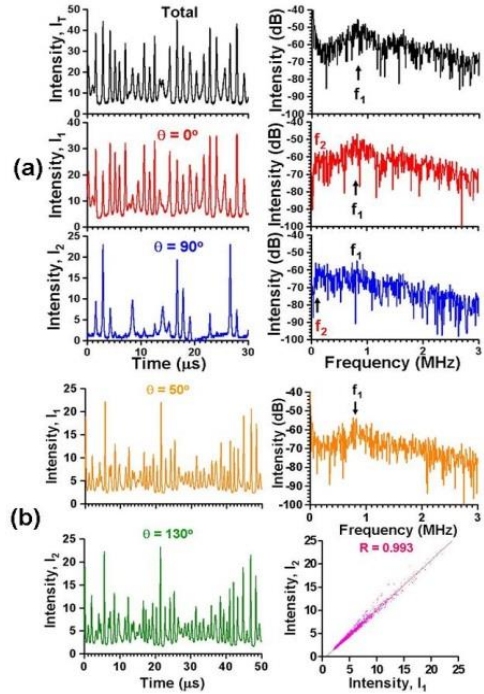

FIG. 4. (a) Chaotic oscillation of DPO eigenmodes. (b) Synchronized chaotic oscillation of a single pair of polarized transverse modes. P = 172 mW.

When complete transverse mode-locking is established, the vector laser behaves as an 'all-in-one' single mode laser and all polarization-resolved chaotic waveforms are automatically synchronized [13]. In the quasi-locked states, only a single pair of polarized transverse fields is found to exhibit chaos synchronization. Moreover, a high degree of chaos synchronization occurred between the part of the total output that did not pass through the polarizer and either of the paired modes. This implies that the chaotic regime inherits the self-organized synchronization of noise-driven relaxation oscillations among a pair of transverse modal outputs and the total output without self-mixing modulations is inherited in the chaotic regimes.

Such a dynamical non-independence of a single pair of transverse modal outputs and the total output in the chaotic regime is inherent to the quasi-locked states; a high-degree of chaos synchronization between a single pair of transverse modal outputs and the total output occurred when one mode was subjected to self-mixing modulation, where part of the polarized output along $+\theta_c$ (or $-\theta_c$) was focused on the rotating cylinder (see the set-up enclosed in the dashed line in Fig. 1(a)). The example shown in Figs. 5(a)-(b) indicates a high degree of chaos synchronization, with a correlation coefficient of R = 0.995. In other words, the modulated mode acts as an information 'sender', while the partner mode acts as a 'receiver' and the chaotic signal is transferred to the partner mode, which is embedded in the transverse vector lasing fields, forming a coherent field together with the modulated mode. Figure 5(c) shows the change in the amplitude correlation coefficient, R, between a pair of transverse vector fields when the polarization direction of the probe beam (i.e., receiver) was varied. This sender-receiver type of synchronization shows potential for a secure metrology system, where a self-mixing signal created by the output beam toward a target [20], which is polarized along $\pm\theta_c$, can be accurately retrieved solely by the output beam polarized along $\mp\theta_c$.

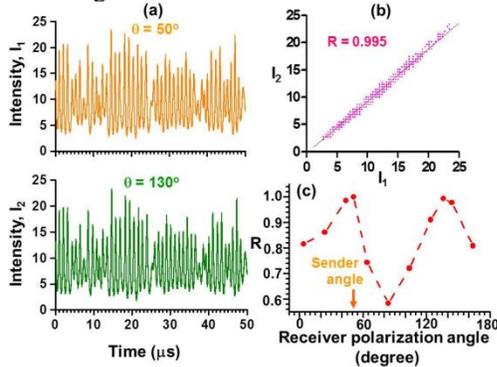

FIG. 5. (a) Synchronized chaos for LDV modulation of one of the paring modes (i.e., the sender) and (b) correlation plots. (c) Dependence of R on polarization direction of the receiver. P = 172 mW.

Essentially the same synchronization between the single pair of transverse modes shown in Figs. 4 and 5 occurred for quasi-locked states among the $IG^e_{2,2} - HG_{0,0}$ as well as the $TEM_{0,1}{}^* - HG_{0,0}$ modes in Fig. 1(b), for both total and one-beam feedback from the rotating cylinder. The results for the $+\theta_c$-polarized beam feedback are shown in Fig. 6.

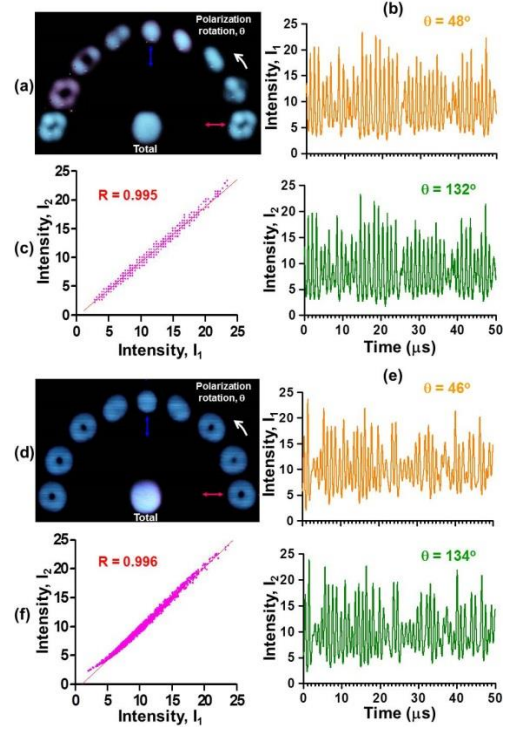

FIG. 6. Synchronized chaos. (a)-(c) $IG^2_{2,2} - HG_{0,0}$, P = 180 mW. (d)-(f) $TEM_{0,1}{}^* - HG_{0,0}$, P = 200 mW.

On the basis of repeated experiments, the critical angles are found to be approximately $\theta_c \cong \arctan(\pm 1/r)$ when the subsidiary peaks are much smaller compared with the dominant peaks. Here, the field components of the orthogonally polarized eigenmodes along $\theta_c$ coincide and the effective modal gain for one mode is almost the same as that for the other mode. In short, such a pair of transverse modes acts just like a single-mode laser through coherent field coupling among modes.

Numerical simulations were conducted for comparison with the experimental results. The simulation used the following two-mode laser equations, which include *incoherent* modal cross-saturation and *coherent* field coupling in the short delay regime, i.e., $\tau_D \ll 1/f_1$.

$$\frac{dN_1}{dt} = \left(\frac{2}{K}\right)\left[\frac{P}{P_{th}} - 1 - N_1 - (1 + 2N_1)(E_1^2 + \beta_{1,2}E_2^2)\right], \quad (1)$$

$$\frac{dN_2}{dt} = \left(\frac{2}{K}\right)\left[\frac{P}{P_{th}} - 1 - N_2 - (1 + 2N_2)(E_2^2 + \beta_{2,1}E_1^2)\right], \quad (2)$$

$$\frac{dE_1}{dt} = g_1 E_1 N_1 + m_1 E_1 \cos\Omega_{D,1} t + \eta E_2 \cos\Psi + \sqrt{2\varepsilon[N_1 + 1]}\,\xi(t), \quad (3)$$

$$\frac{dE_2}{dt} = g_2 E_2 N_2 + m_2 E_2 \cos\Omega_{D,2} t + \eta E_1 \cos\Psi + \sqrt{2\varepsilon[N_2+1]}\,\xi(t), \qquad (4)$$

$$\frac{d\Psi}{dt} = -\eta\left(\frac{E_2}{E_1} + \frac{E_1}{E_2}\right)\sin\Psi \qquad (5)$$

Here, $N_i$ is the modal excess population inversion, $E_i$ is the modal field amplitude, $\beta_{i,j}$ is the cross-saturation coefficient, $\eta$ is the coupling coefficient between modal fields, $g_i$ is the relative modal gain coefficient, $\Omega_{D,i} = 2\pi f_{D,i}/\kappa$ ($\kappa = 1/2\tau_p$: damping rate of the cavity) is the normalized Doppler-shift frequencies, $m_i$ is the amplitude feedback rate, $\Psi$ is the modal phase difference, $\varepsilon$ is the spontaneous emission coefficient, and $\xi(t)$ is the Gaussian white noise with zero mean and $<\xi(t)\xi(t')> = \delta(t-t')$. $K = \tau/\tau_p$ is the fluorescence-to-photon lifetime ratio. Time is normalized by $2\tau_p$.

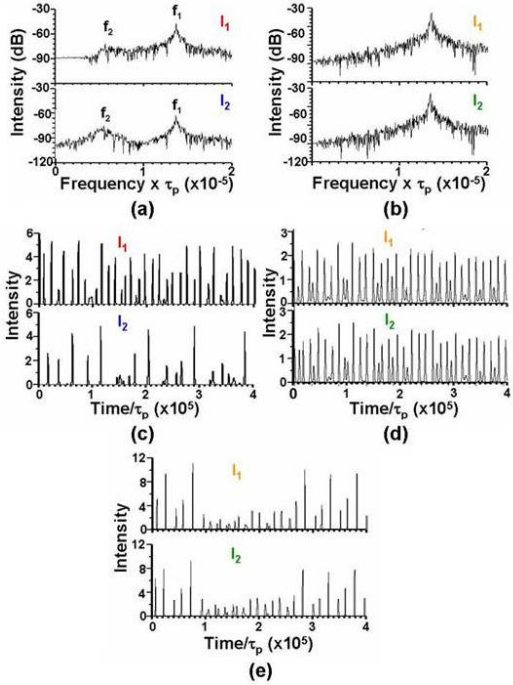

FIG. 7. Numerical results. (a), (c) $g_1 = 1$, $g_2 = 0.70$, $\beta_{1,2} = 0.46$, $\beta_{2,1} = 0.75$, $\eta = 0.001$ (b)-(d) $g_1 = g_2 = 0.9$, $\beta_{1,2} = \beta_{2,1} = 0.6$, $\eta = 0.005$. (a), (b) Free-running, $m_1 = m_2 = 0$. (c), (d) Total modulation, $m_1 = m_2 = 0.002$. (e) One-mode modulation, $m_1 = 0.002$, $m_2 = 0$.

The numerical results are shown in Fig. 7, where w = 1.53, $K = 3.37 \times 10^5$, $f_D = 2.37 \times 10^5$ and $\varepsilon = 8.3 \times 10^{-10}$. As for the pair of transverse eigenmodes, the free-running modal power spectra exhibit an $f_2$-peak representing antiphase dynamics, as shown in Fig. 7(a), whereas this peak disappears for the single pair of modes of the common gain and an increased field coupling through synchronization, as shown in Fig. 7(b).

When the self-mixing modulation is applied to the total ($m_1 = m_2$) or one mode ($m_2 = 0$), a high degree of chaos synchronization occurs between the single pair of modes of the common gain in both cases, as shown in Figs. 7(d)-(e), while synchronization fails for the pair of transverse eigenmodes as shown in Fig. 7(c).

In summary, synchronized chaotic relaxation oscillations were found to occur between a single pair of polarized transverse mode fields in the quasi-locked states of coupled orthogonally polarized transverse modes in a thin-slice solid-state laser subjected to self-mixing modulations. The sender-receiver relationship of the chaotic signals hidden in the lasing polarization vector fields occurs between a single pair of transverse fields. This relationship inherits the dynamical non-independence in noise-driven modal intensity fluctuations. The experimental results were well reproduced by numerical simulations of a model of two-mode lasers subjected to self-mixing modulations, which included incoherent cross-saturation and coherent coupling.